\documentclass[twocolumn,pra,superscriptaddress]{revtex4}
\usepackage{amsfonts}
\usepackage{amssymb}
\usepackage{amsmath}
\usepackage{epsfig}
\usepackage{color}
\usepackage{graphics, graphicx}
\usepackage{bbold}
\usepackage{psfrag}
\usepackage{mathcomp}
\usepackage{subfigure}
\usepackage{verbatim}
\usepackage{float}
\usepackage{graphicx}
\usepackage[colorlinks,citecolor=blue]{hyperref}
\makeatletter

\newcommand{\Rmnum}[1]{\expandafter\@slowromancap\romannumeral #1@}
\makeatother

\begin{document}

\title{Observation of higher-order time-dislocation topological modes}

\author{Jia-Hui Zhang}
\affiliation{State Key Laboratory of Quantum Optics and Quantum Optics Devices, Institute of Laser Spectroscopy, Shanxi University, Taiyuan, Shanxi 030006, China}

\author{Feng Mei}
\email{meifeng@sxu.edu.cn}
\affiliation{State Key Laboratory of Quantum Optics and Quantum Optics Devices, Institute of Laser Spectroscopy, Shanxi University, Taiyuan, Shanxi 030006, China}
\affiliation{Collaborative Innovation Center of Extreme Optics, Shanxi University, Taiyuan, Shanxi 030006, China}

\author{Yi Li}
\affiliation{State Key Laboratory of Quantum Optics and Quantum Optics Devices, Institute of Laser Spectroscopy, Shanxi University, Taiyuan, Shanxi 030006, China}

\author{Ching Hua Lee}
\email{phylch@nus.edu.sg}
\affiliation{Department of Physics, National University of Singapore, Singapore 117551, Republic of Singapore}

\author{Jie Ma}
\affiliation{State Key Laboratory of Quantum Optics and Quantum Optics Devices, Institute of Laser Spectroscopy, Shanxi University, Taiyuan, Shanxi 030006, China}
\affiliation{Collaborative Innovation Center of Extreme Optics, Shanxi University, Taiyuan, Shanxi 030006, China}

\author{Liantuan Xiao}
\affiliation{State Key Laboratory of Quantum Optics and Quantum Optics Devices, Institute of Laser Spectroscopy, Shanxi University, Taiyuan, Shanxi 030006, China}
\affiliation{Collaborative Innovation Center of Extreme Optics, Shanxi University, Taiyuan, Shanxi 030006, China}

\author{Suotang Jia}
\affiliation{State Key Laboratory of Quantum Optics and Quantum Optics Devices, Institute of Laser Spectroscopy, Shanxi University, Taiyuan, Shanxi 030006, China}
\affiliation{Collaborative Innovation Center of Extreme Optics, Shanxi University, Taiyuan, Shanxi 030006, China}

\begin{abstract}
 Topological dislocation modes resulting from the interplay between spatial dislocations and momentum-space topology have recently attracted significant interest. Here, we theoretically and experimentally demonstrate time-dislocation topological modes which are induced by the interplay between temporal dislocations and Floquet-band topology. By utilizing an extra physical dimension to represent the frequency-space lattice, we implement a two-dimensional Floquet higher-order topological phase and observe time-dislocation induced $\pi$-mode topological corner modes in a three-dimensional circuit metamaterial. Intriguingly, the realized time-dislocation topological modes exhibit spatial localization at the temporal dislocation, despite homogeneous in-plane lattice couplings across it. Our study opens a new avenue to explore the topological phenomena enabled by the interplay between real-space, time-space and momentum-space topology.
\end{abstract}

\maketitle

\section{Introduction}

As critical subsystems that straddle different topological phases, topological interfaces are host to exotic phenomena ranging from robust chiral anomalies to percolating edge states~\cite{Lu2014,Ozawa2019}. Particularly in higher dimensions, such interfaces become especially rich due to the directional interplay between different localization mechanisms, as epitomized by higher-order~\cite{BBH,2017Brouwer,2017Fang,2018Neupert,2018Lee,2018serra,2018Bahl,2018Rechtsman,2019Zhang,2019Khanikaev,2019Jiang,qi2020acoustic,2021Xie,luo2021observation,Chen2021,koh2023observation,shang2024observation} and hybrid skin-topological states~\cite{lee2019hybrid,yao2018edge,zhang2022review,lin2023topological,lei2024activating,zhu2024brief}. Of late, the prospect of time-periodic (Floquet) driving has introduced yet another new level of complexity by opening up the temporal dimension~\cite{2013Rechtsman,2016Zhang,2016Alu,2016Zhu,maczewsky2017observation,2017Thomson,lee2018floquet,2019Zhen,yang2020photonic,YangZJ2022,Zhu2022time}.

Drawing inspiration from topological phenomena and singularities arising from spatial lattice dislocation~\cite{JiangJH2023review,Fan2018,Jiang2018,Xue2021,LiuZ2022,Mordechai2022,Bahl2022,lin2022topological}, it naturally prompts the question of whether intriguing new physics may also emerge from dislocations in temporal space~\cite{Wang2017a,Wang2017b}. Time-varying materials have recently garnered significant attention due to their rich physics and potential for novel functionalities~\cite{lustig2018topological,lyubarov2022amplified,Alu2023a,Alu2023b,ye2023reconfigurable,Yan2024}. Existing phenomena induced by temporal variations or interfaces, such as temporal photonic crystals\cite{lustig2018topological}, metasurfaces\cite{cai2021dynamically}, temporal reflection and refraction~\cite{Alu2023a,Alu2023b,ye2023reconfigurable,Yan2024}, have already found great promise in new signal processing applications, transcending the scope of space-varying materials in many aspects. However, interfaces between subsystems at different times have remained relatively elusive, partly due to their perceived experimental inaccessibility.

In this work, we for the first time theoretically and experimentally report observation of higher-order Floquet topological corner modes (TCMs) at the interface between subsystems effectively occurring at different times. This is achieved by representing our two-dimensional (2D) higher-order topological system in the frequency space, such that multiple copies of it are stacked along an additional frequency dimension. Notably, each 2D layer is translation invariant across the temporal boundary, even though the TCMs are spatially localized around it. Our resultant 3D lattice is physically realized in an electrical circuit metamaterial, whose freedom in node connectivity allows for versatile implementation of its Floquet driving terms as inter-layer couplings. While electrical circuit arrays have been previously harnessed to realize static nodal and higher-order topological phenomena~\cite{ZhangX2019,LiuZ2020,ZhangS2020,ZhangX2021,ZhangX2022a,ZhangX2022b,Yan2022square,wang2023realization,zhang2023electrical,LiYi2023,shang2024observation}, this is the first time it has been used to realize anomalous Floquet modes.

\begin{figure*}[htbp]
	\centering
	\includegraphics[scale=0.45]{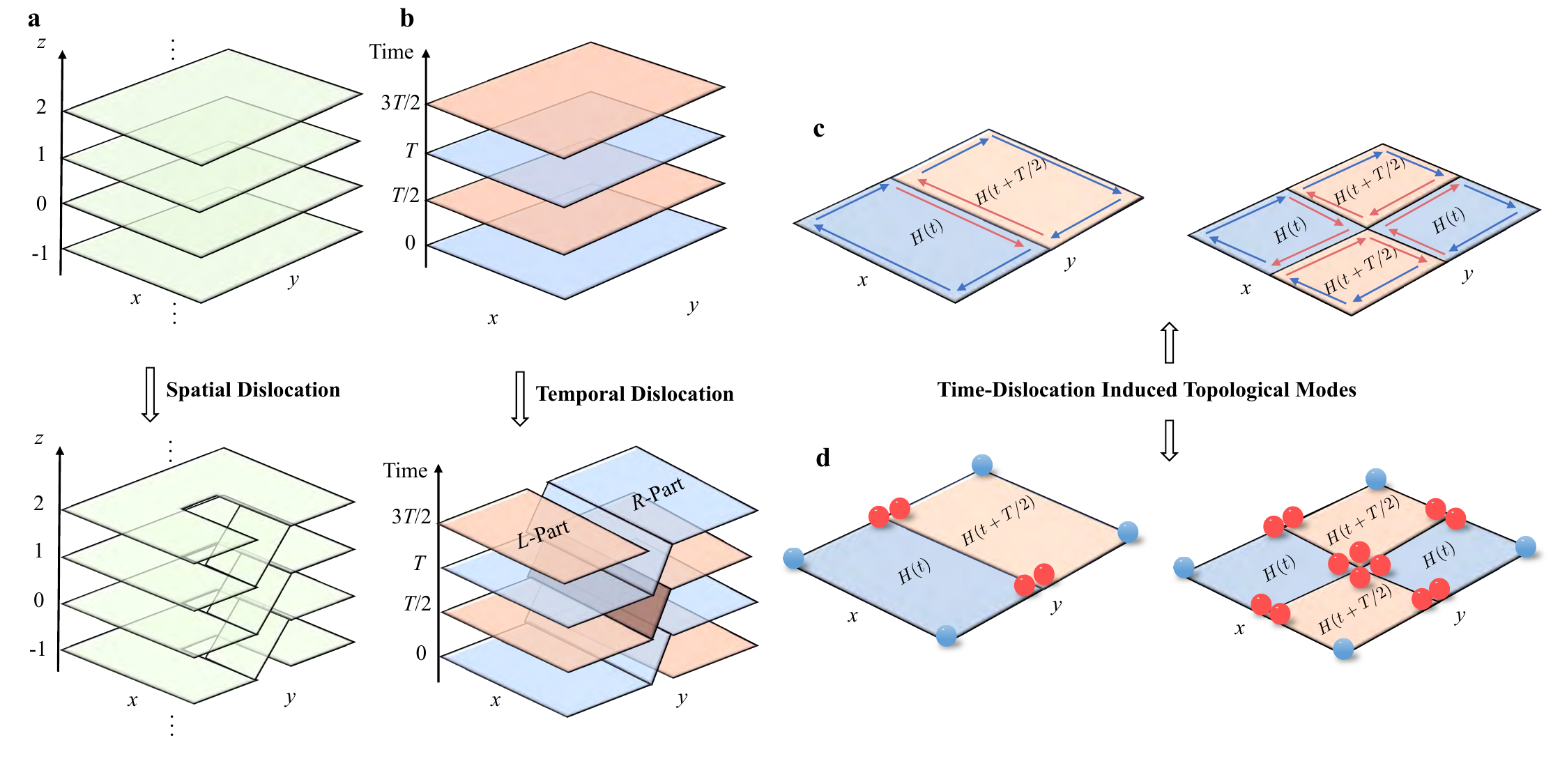}
	\caption{\textbf{Spatial and temporal dislocations}. (a) A spatial lattice dislocation such as the screw dislocation breaks translation symmetry in the $z$-spatial direction. (b) A temporal dislocation can be generated by cutting-and-gluing different stages of a periodically driven system, and breaks time translation invariance. Temporal dislocations can give rise to Floquet topological (c) edge and (d) corner modes within the first- and second-order topological phases, respectively, even though the system remains spatially translation invariant. The red and blue balls (skinny arrows) represent the TCMs (TEMs) induced by the temporal dislocations and spatial boundaries, respectively.}
	\label{Fig1}
\end{figure*}

\begin{figure*}[htbp]
\centering
\includegraphics[scale=0.6]{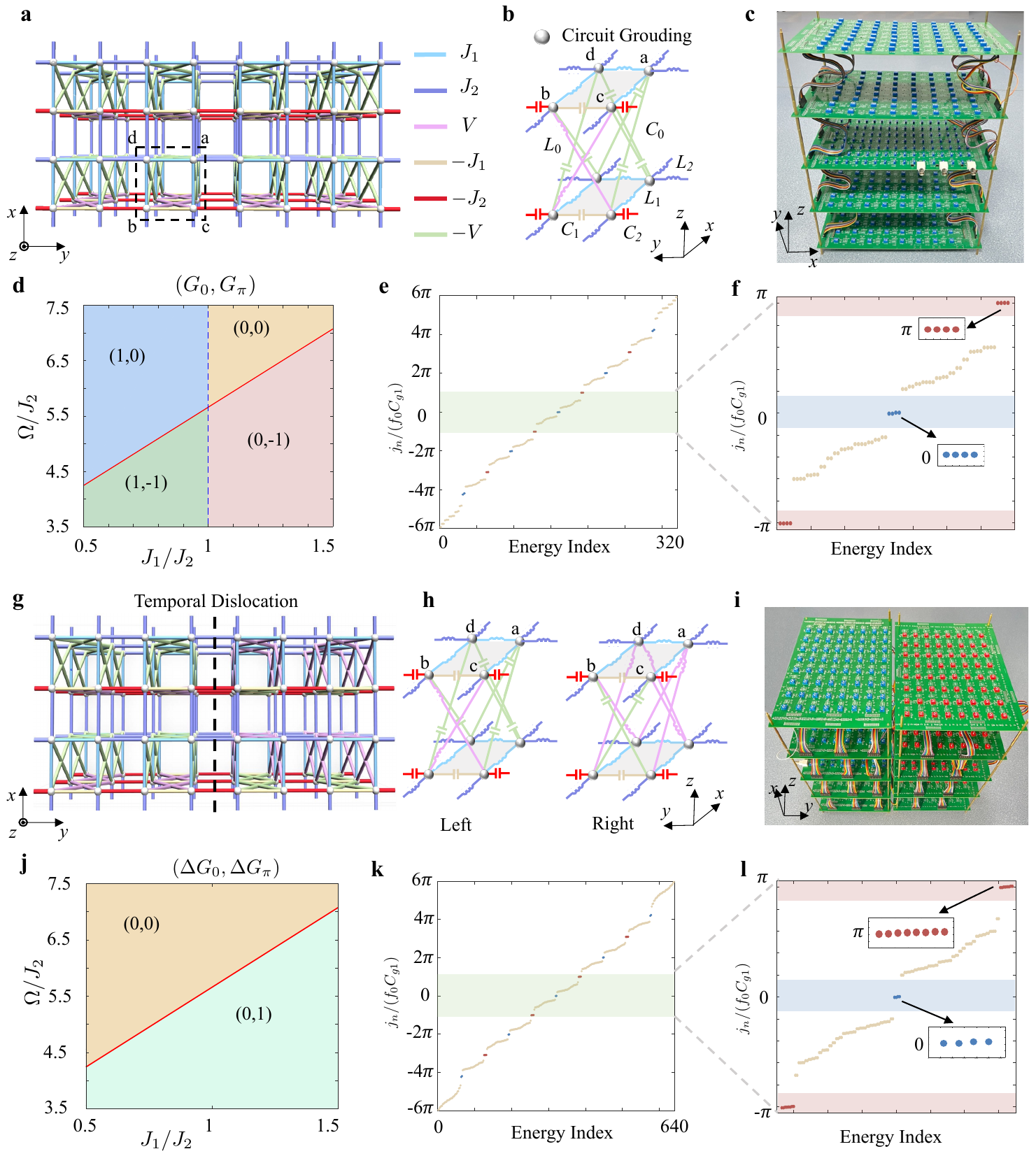}
\caption{\textbf{Time-dislocation induced Floquet higher-order topological modes}. (a) Three-dimensional lattice model representing $H(t)$ in the frequency space. The $z$ direction simulates the frequency dimension, with different $z$ layers representing distinct frequency copies. (b) Circuit diagram for the inter- and intra-layer couplings. (c) Photograph of printed circuit board for implementing Eq. (\ref{JcL}). (d) Floquet mirror winding numbers $(G_0,G_{\pi})$ as a function of $J_1/J_2$ and $\Omega/J_2$ for $V=J_1$, showing topologically nontrivial phases in much of parameter space except for $(0,0)$. (e) Simulated admittance spectrum at the resonant frequency $\omega_0=2\pi\times 1.52$ MHz ($f_0=\omega_0/2\pi$) for the circuit with $J_1/J_2=0.5$ and $\Omega/J_2=2.27$, exhibiting in-gap Floquet topological modes (red and blue points) protected by $(G_0=1,G_{\pi}=1)$. (f) Zoomed-in Floquet spectrum of the real-space analogue of the first quasienergy zone, revealing 4-fold degeneracy in the topological modes. (g) Same as (a) except the right half representing the frequency-space of $H(t+T/2)$, effectively creating a temporal dislocation interface in the middle (dashed line). (h) Circuit diagram for the inter- and intra-layer couplings in the left and right sides of the temporal dislocation, and (i) the fabricated printed circuit board for implementing Eq. (S13) in the Supplementary Note 5. (j-l) Same as (d-f) but for dislocation topological invariants $(\Delta G_0, \Delta G_{\pi})$ and the admittance spectra in the $(\Delta G_0=0, \Delta G_{\pi}=1)$ phase.}
\label{Fig2}
\end{figure*}

\bigskip
\section{Time-dislocation induced higher-order topological modes}

Unlike well-known lattice defects such as screw dislocations (Fig. \ref{Fig1}(a)) which break spatial translation invariance in the $z$ direction, temporal dislocations are defects that disrupt translation invariance along the time dimension. In periodically driven Floquet systems $H(t)$, temporal dislocations occur at the interfaces between different stages of the Floquet drive (Fig. \ref{Fig1}(b)), such as between $H(t)$ and $H(t+T/2)$ as in this work. As shown in Figs. \ref{Fig1}(c,d), temporal dislocations in Floquet topological lattices can lead to the emergence of Floquet topological edge modes (TEMs) and TCMs (see Supplementary Notes 2 and 3 for more details). This work specifically delves into time-dislocation induced TCMs in Floquet higher-order topological phases (HOTPs).

For definiteness, we consider the Benalcazar-Bernevig-Hughes (BBH) model~\cite{BBH} under periodic driving, described by the following Hamiltonian
\begin{align}
H(t)=&(J_1(t)+J_2\cos k_x)\Gamma_1-J_2\sin k_x\Gamma_4 \nonumber \\
	-&(J_1(t)+J_2\cos k_y)\Gamma_3-J_2\sin k_y\Gamma_2
\label{Ht}
\end{align}
where the periodic driving is applied on the intra-cell couplings via $J_1(t)=J_1-2V\cos(\Omega t)$, with $V$ and $\Omega$ being the driving amplitude and frequency, respectively. $\Gamma_1=\tau_x\sigma_0$ and $\Gamma_{2,3,4}=\tau_y\sigma_{x,y,z}$ are the Dirac matrices defined in the sublattice space, and $J_{1,2}$ represents the intra- and inter-cell couplings. In the absence of periodic driving, the above Hamiltonian is reduced to the static BBH model, featuring nontrivial (trivial) static HOTPs for $J_1<J_2$ ($J_1>J_2$), characterized by the mirror winding number $\nu=1$ ($\nu=0$)\cite{BBH}.

Distinct from the static BBH model, the periodically driven BBH model features periodic quasienergy spectra and both 0- and $\pi$-mode gaps~\cite{Bomantara2019,Huang2020}. To identify its bulk topology, we introduce Floquet mirror winding numbers $G_0$ and $G_\pi$, which are defined in terms of the Floquet operators (see the Supplementary Note 1 for the definition). The nontrivial (nonzero) values of $G_0$ and $G_{\pi}$ respectively identify the 0- and $\pi$-mode gaps as topologically nontrivial. The corresponding topological phase diagram, depicted in Fig. \ref{Fig2}(d), is derived as a function of the ratio $J_1/J_2$ and the driving frequency $\Omega$. As shown, in sharp contrast to the static BBH model, which hosts two kinds of HOTPs, the periodically driven BBH model exhibits four distinctive nontrivial Floquet HOTPs, distinguished by the Floquet topological invariants $(G_0=0,1, G_{\pi}=0,-1)$. Notably, as $J_1 >J_2$, the system even remains topologically nontrivial. Contrary to the static BBH model, this case is topologically trivial. The transitions between different Floquet HOTPs are determined by the closing of 0- or $\pi$-mode gaps, as indicated by the dashed or solid lines.

Although it is in-principle possible to realize the real-time modulated coupling in photonic~\cite{2013Rechtsman}, acoustic~\cite{chen2024transient}, electrical circuits~\cite{Pumping2024} or ultracold atomic~\cite{ChenS2023} platforms, for this work, we physically implement the temporal driving as physical couplings along an additional frequency dimension. This not only facilitates the implementation of temporal dislocations, but also allows for the direct observation of Floquet time-dislocation TCMs through steady-state impedance measurements.

The Floquet states for a periodically driven Hamiltonian is most clearly expressed in the frequency space~\cite{Floquetcircuit}. As given in Eq. (\ref{Ht}), the periodically driven BBH model contains a single harmonic driving,
\begin{equation}
H(t) = H_{\text{BBH}} + V_de^{i\Omega t} + V_d^{\dagger}e^{-i\Omega t}
\label{Hr}
\end{equation}
where $V_d=V(\Gamma_3-\Gamma_1)$ and $H_{\text{BBH}}=H(V=0)$.  After transferring into the frequency space, the above 2D lattice Hamiltonian takes a block tridiagonal form (see the Method section) and acquires a third frequency dimension $z$, described by a 3D Hamiltonian
\begin{equation}
H=\sum_{l_z}(H_{\text{BBH}}+z\Omega)\otimes c^{\dag}_{l_z}c_{l_z}+V_d\otimes c^{\dag}_{l_z}c_{l_{z+1}}+\text{H.c.},
\label{Homega}
\end{equation}
where $l_z$ denotes the layer degree of freedom along the frequency space dimension. As depicted in Figs. \ref{Fig2}(a-c), we utilize a circuit metamaterial array to implement the 3D lattice Hamiltonian, where the circuits aligned along the $z$ direction emulate the frequency space~\cite{rudner2020floquet}. Within each $z$ layer ($x$-$y$ plane), the desired positive and negative intra- and inter-cell couplings are implemented by the inductors $L_1,L_2$ and capacitors $C_1,C_2$,
together realizing the BBH model. The layer-specific on-site energy shifts are engineered by carefully designing the grounding terms in each layer (see the Supplementary Note 4). The couplings between nearest-neighbour layers are achieved by the inductor $L_{0}$ and the capacitor $C_{0}$. At the resonant frequency, $\omega_0=1/\sqrt{L_1C_1}=1/\sqrt{L_2C_2}=1/\sqrt{L_0C_0}$, the circuit Laplacian reads
\begin{align}
	J(\omega_0)=&\textstyle\sum_{l_z} (J_0(\omega_0)+\Omega_z) c^{\dag}_{l_z}c_{l_z}
	+V(\omega_0)c^{\dag}_{l_z}c_{l_{z+1}}+\text{H.c}
\label{JcL}
\end{align}
where $\Omega_z=i\sqrt{{C_0}/{L_0}}z\Omega$, $V(\omega_0)=i\sqrt{{C_0}/{L_0}}(\Gamma_3-\Gamma_1)$ and $J_0(\omega_0)=i\sqrt{{C_0}/{L_0}}H_{\text{BBH}}$, with $C_{1,2}=J_{1,2}C_0$ and $L_{1,2}=L_0/J_{1,2}$.

Due to topological bulk-boundary correspondence, the values of $(G_0, G_{\pi})$ respectively determine the numbers of 0- and $\pi$-mode (or $-\pi$-mode) TCMs in the corners of this lattice. To confirm this, as exemplified in Fig. \ref{Fig2}(e) for the phase $(G_0=1, G_{\pi}=-1)$, we numerically simulate the admittance spectrum of the circuit Laplacian in Eq. (\ref{JcL}) under open boundary conditions (OBCs). In the simulations, the truncation of the frequency lattice layers to $-2\leq z \leq 2$ negligibly affects the results. As illustrated, the corresponding admittance spectrum simulate the quasienergy spectrum of the periodically driven BBH model. Given the localized nature of the eigenstates, as shown in Fig. \ref{Fig2}(f), the spectrum of the truncated Hamiltonian remains a reliable approximation to the exact results within the real-space analogue of the first quasienergy zone (even the second one)~\cite{rudner2020floquet}. Outside of the first and second quasienergy zones, the translational invariance along the frequency dimension is broken due to finite-size truncation. It is clearly observed that, in addition to the four TCMs (blue points) emerged in the 0-mode ($\pm2\pi$-mode) gap, there are also four TCMs (red points) in the $\pm\pi$-mode ($\pm3\pi$-mode) gaps, which are unique to Floquet HOTPs. 

Now, we describe how temporal dislocations are implemented in our physical 3D metamaterial setup. According to Fig. \ref{Fig1}(b), the left and right sections of the system are governed by $H(t)$ and $H(t+ T/2)$, respectively. Upon mapping to frequency space, this temporal dislocation appears as an interface  in the inter-layer couplings along the $z$ (frequency) direction (Figs. \ref{Fig2}(g,i)). Notably, the intra-layer couplings within the $x$-$y$ plane remains untouched by the temporal interface, and are perfectly identical on either side of it, even though the interface TCMs would accumulate within the $x$-$y$ plane. This interface is realized by appropriately designing $z$-direction circuit elements (Fig. \ref{Fig2}(h)). In particular, we find that the frequency space of $H(t+T/2)$ can be implemented by designing the right-side $z$-direction inter-layer couplings $V(\omega_0)=-i\sqrt{{C_0}/{L_0}}(\Gamma_3-\Gamma_1)$ (see the Supplementary Note 5 for more details).

The topological feature for the temporal dislocation is characterized by the Floquet topological invariants of its two sides, i.e., $G^{L,R}_{0,\pi}$. Specifically, we define two dislocation topological invariants
\begin{equation}
\Delta G_{0,\pi}=\frac{G^{R}_{0,\pi}-G^{L}_{0,\pi}}{2},
\end{equation}
which respectively determine the numbers of 0- and $\pi$-mode TCMs at the temporal dislocation interface. In terms of ($\Delta G_0, \Delta G_{\pi}$), Fig. \ref{Fig2}(j) illustrates that the system, situated within the same parameter space as Fig. \ref{Fig2}(d), exhibits nontrivial topological dislocations, characterized by ($\Delta G_0=0,\Delta G_{\pi}=1$), even for $J_1>J_2$. As a concrete example, we showcase the TCMs in Figs. \ref{Fig2}(k, l) for $(G^{L}_{0}=1, G^{L}_{\pi}=-1)$ and $(G^{R}_{0}=1, G^{R}_{\pi}=1)$. According to ($\Delta G_0=0, \Delta G_{\pi}=1$), there would emerge time-dislocation induced $\pi$-mode TCMs. The simulated admittance spectra show that, the spectra in the first and second quasienergy zones provide a good approximation to the exact quasienergy spectrum. Alongside the four TCMs (blue points) in the 0-mode ($\pm2\pi$-mode) gaps, there are eight $\pm\pi$-mode TCMs (red points) in the $\pm\pi$-mode ($\pm3\pi$-mode) gaps. As manifested by their density distributions in Fig. \ref{Fig4}, four of the eight $\pm\pi$-mode TCMs are the time-dislocation induced TCMs.

\begin{figure*}[htbp]
	\centering
	\includegraphics[scale=0.62]{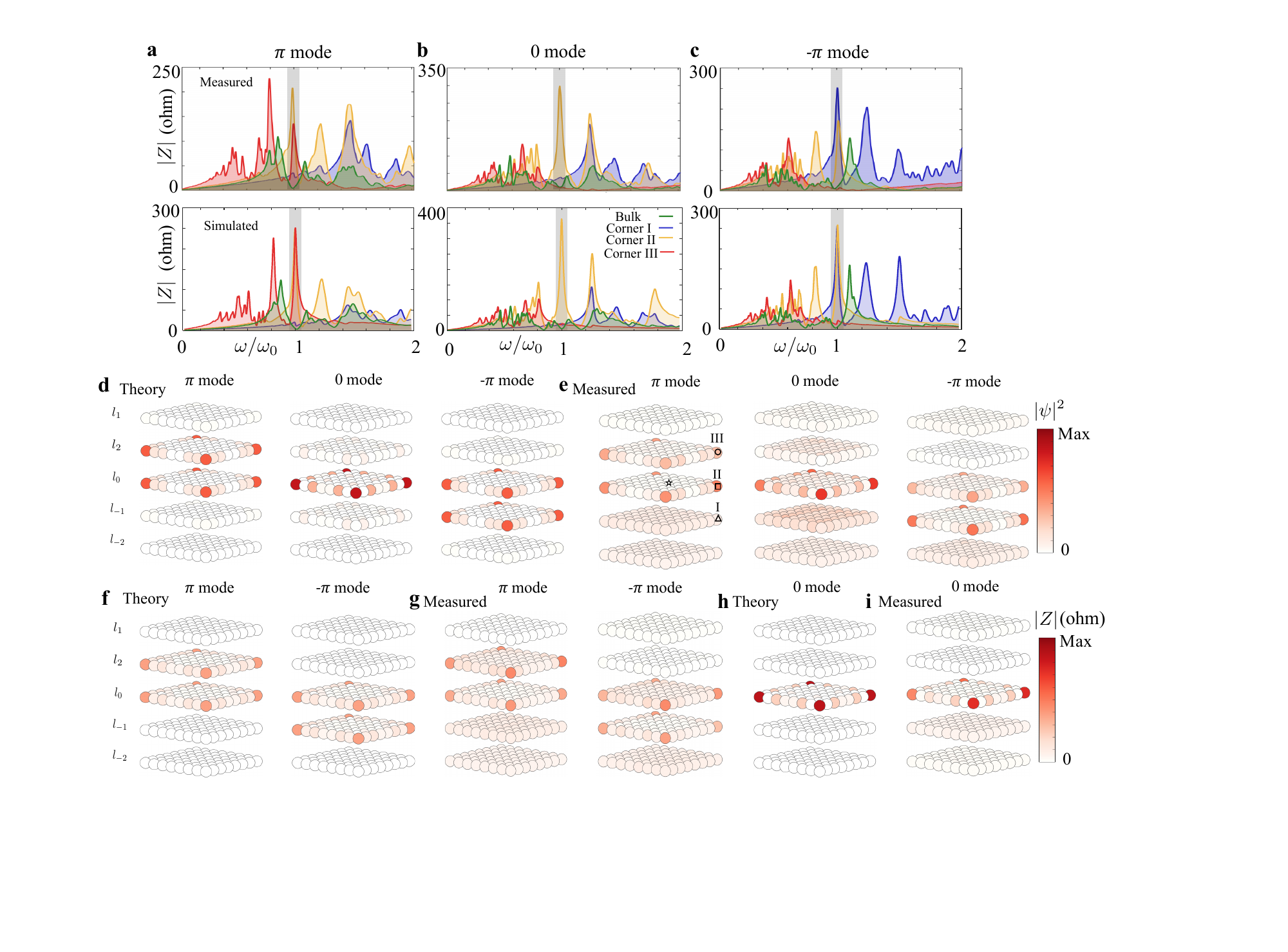}
	\caption{\textbf{Observation of Floquet higher-order topological modes}. (a-c) Measured and simulated impedances over the ground for $(G_0=1,G_{\pi}=-1)$ versus the driving frequency at the bulk and corner I, II and III circuit nodes (marked separately by the star, triangle, square and circle in (e)), respectively corresponding to excite the $\pi$, $0$ and $-\pi$ modes. (d,e) Theoretical density distributions for the emerged in-gap modes and measured impedance distributions in all circuit nodes at the resonant frequency $\omega_0=2\pi\times1.52$ MHz. (f,g) and (h,i) Same as (d,e) but respectively for $(G_0=0,G_{\pi}=-1)$ and $(G_0=1,G_{\pi}=0)$. The parameters are $C_0=1.1\ \text{nF}$, $L_0=10\ \mu\text{H}$, (a-e,h,i) $C_1=1.1\ \text{nF}$, $C_2=2.2\ \text{nF}$, $L_1=10\ \mu\text{H}$, $L_2=5\ \mu\text{H}$, and (f,g) $C_1=2.2\ \text{nF}$, $C_2=1.1\ \text{nF}$, $L_1=5\ \mu\text{H}$, $L_2=10\ \mu\text{H}$. The circuit parameters for the grounding and on-site energy shifts are presented in the Supplementary Note 4.}
	\label{Fig3}
\end{figure*}
\begin{figure*}[htbp]
	\centering
	\includegraphics[scale=0.63]{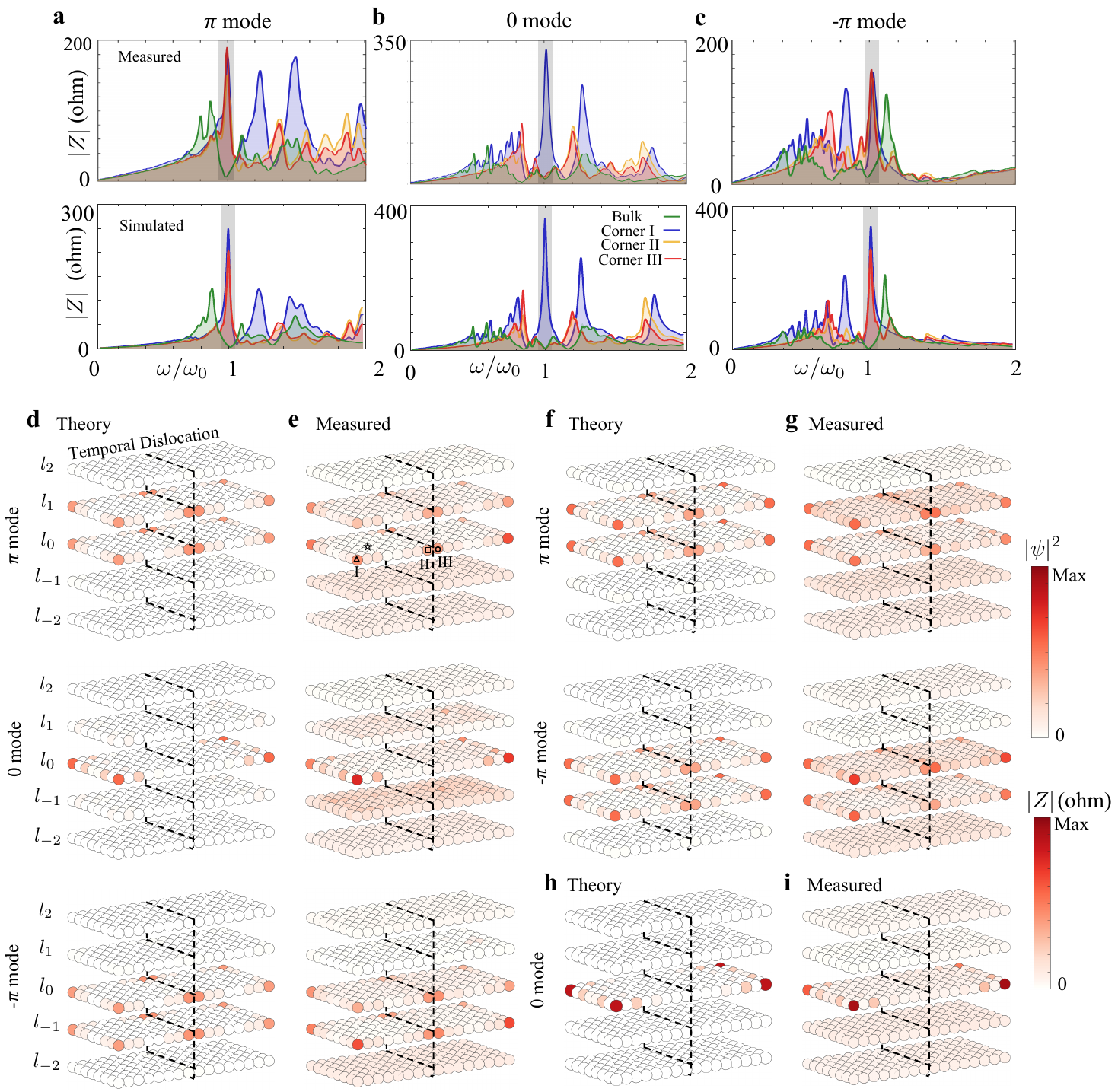}
	\caption{\textbf{Observation of time-dislocation induced Floquet higher-order topological modes}. (a-c) Measured and simulated impedances over the ground for $(G^{L}_{0}=1, G^{L}_{\pi}=-1, G^{R}_{0}=1, G^{R}_{\pi}=1)$ versus the driving frequency at the bulk and corner I, II and III circuit nodes (marked separately by the star, triangle, square and circle in (e)), respectively corresponding to resonantly couple the $\pi$, $0$ and $-\pi$ modes. (d,e) Theoretical density distributions for the emerged in-gap modes and measured impedance distributions in all circuit nodes at the resonant frequency $\omega_0=2\pi\times1.52$ MHz. (f,g) and (h,i) Same as (d,e) but respectively for $(G^{L}_{0}=0, G^{L}_{\pi}=-1, G^{R}_{0}=0, G^{R}_{\pi}=1)$ and $(G^{L}_{0}=1, G^{L}_{\pi}=0, G^{R}_{0}=1, G^{R}_{\pi}=0)$. The parameters are $C_0=1.1\text{nF}$, $L_0=10\ \mu\text{H}$, (a-e,h,i) $C_1=1.1\ \text{nF}$, $C_2=2.2\ \text{nF}$, $L_1=10\ \mu\text{H}$, $L_2=5\ \mu\text{H}$, and (f,g) $C_1=2.2\ \text{nF}$, $C_2=1.1\ \text{nF}$, $L_1=5\ \mu\text{H}$, $L_2=10\ \mu\text{H}$. The circuit parameters for the grounding and on-site energy shifts are presented in the Supplementary Note 5.}
  \label{Fig4}
\end{figure*}

\section{Experimental observation of Floquet higher-order topological modes}

Experimentally, as depicted in Fig. \ref{Fig2}(c), we implement the frequency space of Floquet BBH model with a 3D electrical circuit metamaterial, comprising layers of $8\times8$ in the $x$-$y$ plane and 5 layers extending along the $z$ direction. To spectrally demonstrate the in-gap modes in Fig. \ref{Fig2}(f) in the $(G_0=1, G_{\pi}=-1)$ phase as TCMs, we drive the corner or bulk circuit nodes at the position $r$ and measure the corresponding impedances over the ground, denoted as $Z_r=\sum_{n}|\psi_{n,r}|^2/j_n$. The measured impedances in Fig. \ref{Fig3}(b), as a function of the driven frequency, exhibit excellent agreement with simulated results. Notably, a maximal impedance peak occurs at the corner II precisely at the resonant frequency $\omega_0$, in stark contrast to the absence of such peaks at the corner I, corner III, and the bulk nodes. This discrepancy serves as unequivocal evidence for the existence of 0-mode TCMs ($j_n=0$) situated at the corners of the layer $l_0$. However, since the driving at the resonant frequency only excites the zero mode, the above method cannot directly detect the $j_n/(f_0C_{g1})=\pm\pi$ modes.

To address this challenge, our strategy is to shift the entire admittance spectrum upwards (or downward) by $\pi$, while ensuring the resonant frequency remains unchanged. In the Supplementary Note 6, we provide the detailed procedures regarding the redesign of the grounding circuits for achieving this objective. This shift results in the $-\pi$ ($\pi$) mode in the admittance spectrum moving downwards (upwards) and transforming into $j_n=0$. Consequently, we can now excite the $\pm\pi$ modes by driving the corner node at the same resonant frequency $\omega_0$. The corresponding measured impedances in Figs. \ref{Fig3}(a,c) clearly indicate that, the maximal impedance peaks appear at the corner II and III (I) nodes for the case of the $\pi$ ($-\pi$) modes excited, manifesting that the $\pi$ ($-\pi$) modes predominantly occupy the corners of the layers $l_0$ and $l_{1}$ ($l_{-1}$).

Fig. \ref{Fig3}(d) presents theoretical calculations of the density distributions of the 0- and $\pm\pi$-mode eigenmodes. As uncovered, the 0{-mode} TCMs predominantly occupy the four corners of the layer $l_0$, while the $\pi$-mode ($-\pi$-mode) TCMs are predominantly found at the corners of the layers $l_0$ and $l_1$ ($l_{-1}$). Likewise, the $2\pi$-mode ($-2\pi$-mode) TCMs are primarily localized at the four corners of the layer $l_1$ ($l_{-1}$), and the $3\pi$-mode ($-3\pi$-mode) TCMs are mainly located at the corners of the layers $l_1$ and $l_2$ ($l_{-1}$ and $l_{-2}$). This experiment aims to detect the 0- and $\pm\pi$-mode TCMs, including the next section addressing temporal dislocation. However, it's important to note that the method can be directly utilized to detect the $\pm2\pi$-mode and $\pm3\pi$-mode TCMs as well.

In Fig. \ref{Fig3}(e), we measure the spatial distribution of the impedances in all circuit nodes at the resonant frequency. As shown, the impedances corresponding to the 0 and $\pm\pi$ TCMs excited, maximally occupy the corners of the $l_0$ layer and the $l_{0,\pm1}$ layers, respectively, agreeing with the theoretical predictions in Fig. \ref{Fig3}(d). Figs.~\ref{Fig3}(f,g) and (h,i) respectively investigate the case for the circuit in the $(G_0=0, G_{\pi}=-1)$ and $(G_0=1, G_{\pi}=0)$ phases. As shown, the corresponding circuit only hosts 0- or $\pm\pi${-mode} TCMs, demonstrated by the measured impedances maximally populating the corners of the $l_0$ layer or the corners of the $l_{0,\pm1}$ layers. By comparing the distinct features of measured TCMs, we comprehensively unveil the topological properties of three distinctive nontrivial Floquet HOTPs.

\section{Experimental observation of time-dislocation induced Floquet higher-order topological modes}

Having demonstrated the Floquet HOTPs, we now introduce a time-dislocation induced interface into the circuit metamaterial. As illustrated in Figs. \ref{Fig2}(i), such a temporal dislocation in the frequency space can be directly implemented by a 3D circuit metamaterial array, containing $8\times16\times5$ circuit nodes. The blue and red circuits respectively implement the frequency space of $H(t)$ and $H(t+T/2)$, with the interface representing the temporal dislocation.

To spectrally identify that the $\pi$-mode TCMs in Fig. \ref{Fig2}(l) includes time-dislocation induced TCMs at the corners of the interface, we drive the corner nodes of both the interface and the whole circuit, and measure the corresponding impedances relative to the ground. The measured impedances depicted in Fig. \ref{Fig4}(b) demonstrate a notable agreement with the simulated outcomes. Specifically, at the resonant frequency, a distinct impedance peak is observed at the corner I node, while no such peak is evident at the corners II and III or at the bulk nodes. This disparity emphasizes the 0{-mode} modes as the TCMs of the layer $l_0$. Similarly, we reconfigure the grounding circuits (see the Supplementary Note 6) for exciting the $\pm\pi$ modes at the same resonant frequency $\omega_0$. Figs. \ref{Fig4}(a,c) illustrates that the impedance peaks emerge at all three corners at the resonant frequency, indicating that the $\pm\pi$ modes correspond to the TCMs of the whole circuit and the TCMs of the interface within the layer $l_0$.

In Fig. \ref{Fig4}(d), the density distributions of the 0- and $\pm\pi$-mode eigenmodes are theoretically calculated for compression. It is found that the 0{-mode} TCMs primarily occupy the four corners of the layer $l_0$, while the $\pi${-mode} ($-\pi$-mode) TCMs mainly populate the corners of the layers $l_0$ and $l_1$ ($l_{-1}$) as well as the corners of the interface within these layers. Similarly, the $2\pi${-mode} ($-2\pi$-mode) TCMs are predominantly localized at the four corners of the layer $l_1$ ($l_{-1}$), and the $3\pi${-mode} ($-3\pi$-mode) TCMs are mainly located at the corners of the layers $l_1$ and $l_2$ ($l_{-1}$ and $l_{-2}$) as well as the corners of the interface within these layers.

We also measure the impedance distributions at all nodes at the resonant frequency (Fig. \ref{Fig4}(e)). The experimental data reveals that when the 0 modes are excited, the impedances primarily concentrate at the corners of the layer $l_0$. In contrast, for the $\pm\pi$ modes, the impedances mainly populate the corners of the layers $l_{0,\pm1}$ and the corners of the interface within these layers, aligning with the theoretical predictions in Fig. \ref{Fig4}(d). Figs. \ref{Fig4}(f,g) further explore the situation for the circuit with $(G^{L}_{0}=0, G^{L}_{\pi}=-1, G^{R}_{0}=0, G^{R}_{\pi}=1)$. It becomes evident that the circuit supports only eight $\pm\pi${-mode} modes and does not host 0{-mode} modes. The measured impedance distribution (Fig.~\ref{Fig4}(g)) confirms this observation, as the $\pm\pi${-mode} modes primarily occupy the corners of the layers $l_{0,\pm1}$ and the corners of the interface within these layers, in according with Fig. \ref{Fig4}(f). The results in Figs.~\ref{Fig4}(h,i) for the circuit with $(G^{L}_{0}=1, G^{L}_{\pi}=0, G^{R}_{0}=1, G^{R}_{\pi}=0)$ show that this circuit hosts only four $0${-mode} modes. As expected, the measured impedances mainly concentrate at the corners of the layer $l_0$. The experimental comparison of these distinctive topological scenarios unequivocally demonstrates the emergence of $\pi${-mode} TCMs induced by the temporal dislocation.

\section{DISCUSSION}

In summary, we have theoretically constructed and experimentally observed Floquet higher-order topological corner modes induced by temporal dislocations, making use of an additional physical dimension to represent the frequency-space lattice. Interestingly, our time-dislocation topological modes are spatially localized in-plane across the temporal interface, despite all in-plane couplings being homogeneous across it. 

Our demonstration would impact both experimental and theoretical studies of topological states. On the experimental side, our frequency-space engineering approach demonstrated in our experiment opens up a new pathway for implementing sophisticated Floquet topological models that are not achievable through direct implementation of periodic driving, particularly since high-frequency modulation lead to excessive component heating. This approach has a wide range of applications not limited to circuit platforms, being applicable also in topological photonic and acoustic platforms. On the theory side, our work has revealed that topological modes could also emerge in a temporal junction, alongside time reflection and refraction~\cite{Alu2023a,Alu2023b,ye2023reconfigurable,Yan2024}, representing novel phenomena induced by temporal interfaces. In all, our study also presents an exciting prospect in demonstrating Floquet topological phenomena involving the intricate interplay between real-space, temporal-space and momentum-space topology.

\section{ACKNOWLEDGEMENTS}
We acknowledge Profs. Jian-Hua Jiang, Yiming Pan and Qingqing Cheng for the positive comments. This work is supported by the by the National Key Research and Development Program of China (Grant No. 2022YFA1404201), National Natural Science Foundation of China (NSFC) (Grant No. 12034012, No. 12074234), Changjiang Scholars and Innovative Research Team in University of Ministry of Education of China (PCSIRT)(IRT\_17R70), Fund for Shanxi 1331 Project Key Subjects Construction, 111 Project (D18001) and Fundamental Research Program of Shanxi Province (Grant No. 202303021223005). CHL acknowledges support from the Ministry of Education (MOE), Singapore (Award number: MOE-T2EP50222-0003).

\bigskip

\end{document}